\documentclass[letterpaper]{article} 
\usepackage[preprint]{aaai2027} 
\usepackage[hyphens]{url} 
\usepackage{graphicx} 
\usepackage{natbib} 
\usepackage{caption} 
\usepackage{algorithm}
\usepackage{algorithmic}
\usepackage{amsfonts}
\usepackage{amssymb}
\usepackage{amsmath}
\usepackage{booktabs}
\usepackage{array}
\usepackage{makecell}
\usepackage{adjustbox}
\usepackage[table]{xcolor}
\usepackage{multirow}
\usepackage{cuted}

\definecolor{mem10}{RGB}{231,247,235}
\definecolor{mem05}{RGB}{254,243,226}
\definecolor{refgray}{RGB}{242,242,242}
\definecolor{oursgray}{RGB}{224,224,224}
\definecolor{bbband}{RGB}{237,239,244}

\newcommand{\near}{\textsuperscript{\scriptsize$\approx$}}

\title{Coverage Matters: MarginMerge for Compressing Multi-Vector Visual Document Retrievers}

\author{
Ailar Mahdizadeh\textsuperscript{\rm 1,2},
Aria Salari\textsuperscript{\rm 3},
Sohail Rajabi\textsuperscript{\rm 3},
Shahriar Mirabbasi\textsuperscript{\rm 1},
Panos Nasiopoulos\textsuperscript{\rm 1},
Alireza Morsali\textsuperscript{\rm 3}
}

\affiliations{
\textsuperscript{\rm 1}University of British Columbia \textsuperscript{\rm 2}Vector Institute \textsuperscript{\rm 3}Global Relay\\
ailar.mahdizadeh@ubc.ca
}

\begin{document}

\maketitle
\begin{strip}
    \centering
    \includegraphics[width=0.9\textwidth]{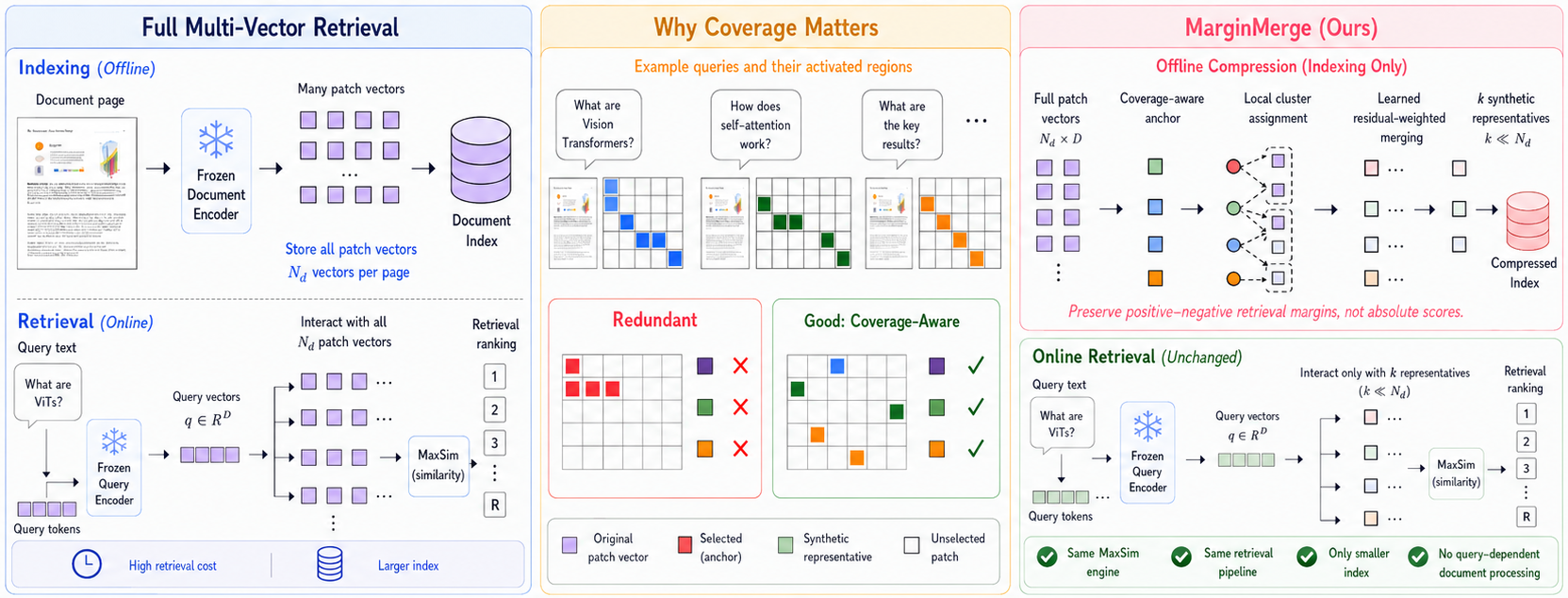}
    \captionof{figure}{
        \textbf{Motivation and overview of MarginMerge.}
MarginMerge compresses multi-vector representations into a compact set of query-relevant representatives while preserving the standard MaxSim retrieval pipeline.
    }
    \label{fig:marginmerge-overview}
\end{strip}

\begin{abstract}
Multi vector visual document retrievers such as ColPali and ColQwen achieve
strong retrieval by storing fine grained patch embeddings, but this produces
large indexes and costly late interaction scoring. We argue that effective
compression should preserve \emph{query relevant coverage}, meaning the diverse
document regions that may become the strongest MaxSim match across queries,
rather than selecting patches independently by salience. This view also
explains why dense rendered pages are easier to compress than natural images. We introduce \textbf{MarginMerge}\footnotemark
\footnotetext{https://github.com/ailarmhz/MarginMerge.git}, a compression method for frozen multi vector
retrievers. It selects coverage aware anchors, clusters document patches, and
uses a lightweight shared network to synthesize one representative per cluster.
Compression is performed once during indexing, while retrieval keeps the
standard MaxSim interface. Across six datasets on both ColQwen2.5 and ColPali, MarginMerge achieves the
highest matched query agnostic average at 5\% and 10\% vector retention.
Compared with the uncompressed index using the same backbone, it preserves
between 97\% and 99\% of average nDCG@5 while reducing stored document vectors
by between 90\% and 95\%. At 5\% retention, it also reduces ranking flips
relative to geometric merging on all six ColQwen2.5 datasets by approximately
41\% on average. The same model transfers to unseen datasets and retention
ratios without retraining.
\end{abstract}

\section{Introduction}
\label{sec:introduction}

Documents impart information through text, figures, tables, typography, and spatial layout. Conventional retrieval systems typically rely on optical character recognition (OCR) or document parsers that may discard visual evidence and require format-specific pre-processing \cite{sun2025unveil,tanaka2025vdocrag}. Visual document retrieval instead operates directly on rendered page images \cite{guo2025nldir,tanaka2025vdocrag}. While Document Screenshot Embedding (DSE) represents each page with a single dense vector \cite{ma2024dse}, ColPali and ColQwen-family retrievers preserve a set of visual patch embeddings and compare them with query-token embeddings through late interaction \cite{faysse2025colpali,ma2025storage,lin2023flmr}. This multi-vector representation allows different query tokens to retrieve localized evidence from different page regions, including text passages, table cells, figures, and layout elements.

This fine-grained representation is central to the effectiveness of ColPali-style retrieval. Given query vectors and document patch vectors, the MaxSim operator assigns each query vector the score of its best-matching document patch, following ColBERT-style late interaction
\cite{khattab2020colbert,santhanam2022colbertv2}. Retaining many contextualized patch vectors therefore preserves localized evidence that may be lost when a page is compressed into a single embedding. However, it also creates a major scalability challenge: a page may contain hundreds or thousands of vectors, so both index size and retrieval cost grow with the number of stored patches \cite{ma2025storage,lee2023xtr,dhulipala2024muvera}. Aggressive compression of this representation can remove precisely the evidence that gives late interaction its advantage. The central challenge is therefore to construct a compact multi-vector representation that preserves the retrieval behavior of the original model.

Figure~\ref{fig:marginmerge-overview} illustrates this challenge and our approach. Full multi-vector retrieval retains all document patch vectors, whereas effective compression must preserve the complementary page regions that may be activated by different queries. To balance broad coverage over complementary regions and compression, MarginMerge constructs a compact set of synthetic representative vectors offline, while leaving the online MaxSim retrieval pipeline unchanged.

Existing compression methods show that visual document representations contain substantial redundancy, but they also expose an unresolved empirical pattern. Similarity-based merging often remains effective at compression levels where patch pruning degrades substantially, and informed importance-based pruning can even underperform uniform random retention \cite{ma2025storage,bolya2023tome}. These findings indicate that compression cannot be treated solely as the identification of individually important patches. What remains unclear is which regions must be preserved, why compressibility varies across document types, and why selecting original patches behaves differently from synthesizing new representatives.

We argue that the missing principle is \emph{query-relevant coverage}. Under MaxSim, each query vector receives its contribution from its best-matching document patch. Although only a small subset of patches may determine the score for one query, different patches may become relevant for different queries. A compact document representation must therefore preserve, or closely approximate, the collection of regions capable of providing strong matches across the query distribution. Compression is consequently a set-coverage problem rather than an independent patch-ranking problem: retaining several highly similar patches may provide less retrieval value than representing several complementary regions.

This perspective also explains why compressibility depends on document content. Dense rendered pages contain repeated glyphs, rows, cells, backgrounds, and layout structures, allowing many patches to substitute for one another. Natural photographs contain more locally distinct visual content and are consequently harder to reduce without losing query-relevant evidence. Scene-text photographs occupy an intermediate regime. We formalize this behavior through the query distribution's \emph{argmax union}, which captures the document patches that may determine MaxSim responses, and through a redundancy statistic that estimates the number of substitutable patches per effective representation direction.

Guided by this analysis, we introduce \textbf{MarginMerge}, a post-hoc method for constructing compact representations from a frozen multi-vector visual retriever. MarginMerge first selects coverage-aware anchors that span complementary query-relevant regions and uses them to partition the original patch vectors into clusters. A lightweight shared network then learns how the patches within each cluster should contribute to a synthetic representative. Rather than reconstructing each document's absolute retrieval score, MarginMerge preserves the positive-negative score margins that determine document ordering. Each page is compressed once during indexing, and the resulting representation can be used in the standard MaxSim interface without query-dependent document processing or changes to the retrieval engine.

Our experiments across rendered documents, scene-text photographs, and natural photographs support this coverage-based account. We show that compressibility varies systematically with representation redundancy and that learned salience-based selectors can fit their training queries while lacking the generalizability to preserve evidence required by unseen queries. We further find that predicting useful patches is insufficient when those patches are retained through concentrated importance ranking; diversity-aware selection more effectively preserves complementary evidence. MarginMerge converts these findings into an effective representation synthesis method that generalizes across datasets and representation budgets while consistently reducing compression-induced ranking changes.

\paragraph{Contributions.}
\begin{itemize}
\item We develop a retrieval-centered account of multi-vector
compressibility based on coverage of the document regions that can become strong matches across a query distribution. This account explains the differing compressibility of rendered documents, scene-text photographs, and natural photographs.

\item We provide a mechanistic analysis of query-agnostic pruning and show why patch salience does not reliably translate into retrieval preservation. Our results identify diverse coverage, rather than concentrated importance, as the key requirement for retrieval quality when selecting original patches.

\item We introduce MarginMerge, a coverage and ranking-aware synthesis method that compresses representations from a frozen visual retriever, preserves the standard MaxSim retrieval interface, and generalizes across datasets and representation budgets.

\end{itemize}

\section{Related Work}
\label{sec:related_work}

\paragraph{Visual document retrieval.}
Visual document retrieval encodes rendered pages to preserve text, images, and
layout without relying fully on OCR. DSE represents each page with one dense
vector \cite{ma2024dse}, while ColPali retains patch embeddings and applies
ColBERT style late interaction \cite{faysse2025colpali,khattab2020colbert}.
ColQwen extends this approach with Qwen vision language backbones, and ViDoRe
evaluates visual retrievers across diverse document collections
\cite{ma2025storage,mace2025vidorev2}. Their fine grained representations
improve retrieval but require storing many vectors per page.

\paragraph{Efficient late interaction retrieval.}
ColBERTv2 compresses individual token vectors, while PLAID accelerates candidate
search and MaxSim scoring
\cite{santhanam2022colbertv2,santhanam2022plaid}. MUVERA approximates multi
vector similarity using fixed dimensional encodings
\cite{dhulipala2024muvera}, and MetaEmbed learns a variable number of meta
tokens \cite{xiao2026metaembed}. These methods modify vector storage, search,
or model representations. MarginMerge instead reduces the number of vectors
produced by a frozen retriever while keeping the standard MaxSim interface.

\paragraph{Visual document representation compression.}
Token merging was introduced for reducing redundant Vision Transformer tokens
\cite{bolya2023tome}. Light ColPali later showed that similarity based merging
is more robust than pruning under aggressive compression and that informed
pruning can underperform random retention \cite{ma2025storage}. Other methods
combine hierarchical reduction and quantization \cite{duong2025hpc}, predict
patch importance \cite{yan2025docpruner}, preserve structural anchors
\cite{liu2026look}. Related work also studies pooling, memory tokens, and
attention guided clustering \cite{qin2026multivector}. These methods show that
multi vector indexes can be compressed, but do not explain why pruning and
merging behave differently across content types. MarginMerge addresses this
through query relevant coverage and learned representative synthesis.

\begin{figure*}[t]
    \centering
    \includegraphics[width=0.9\textwidth]{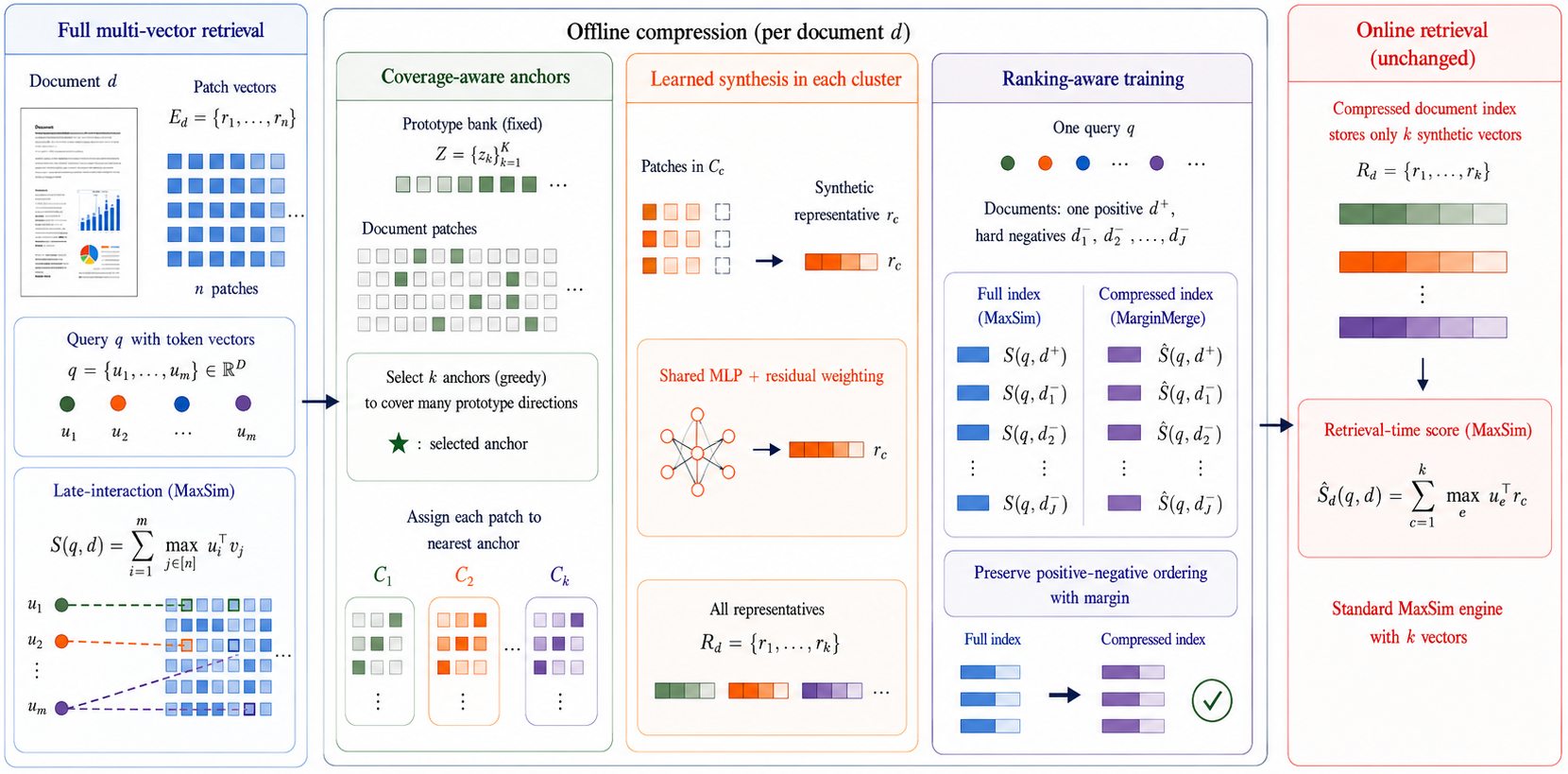}
    \caption{
        \textbf{Overview of MarginMerge.}
The offline pipeline selects coverage-aware anchors and learns one representative per cluster. Ranking-margin distillation preserves full-index ordering, while retrieval uses standard MaxSim over only the compressed representatives.
    }
    \label{fig:marginmerge}
\end{figure*}

\section{Method}
\label{sec:method}

\subsection{Problem Formulation}
\label{sec:problem}




Let a query $q$ and a document $d$ be represented by collections of
$\ell_2$-normalized embedding vectors
\[
E_q = \{\mathbf{u}_1,\ldots,\mathbf{u}_m\},
\qquad
E_d = \{\mathbf{v}_1,\ldots,\mathbf{v}_n\},
\]
where $\mathbf{u}_i$ and $\mathbf{v}_j$ are $D$-dimensional query-token and
document-patch embeddings, respectively. Late-interaction retrievers score a
query--document pair using MaxSim:
\begin{equation}
S(q,d)
=
\sum_{i=1}^{m}
\max_{j \in [n]}
\mathbf{u}_i^{\top}\mathbf{v}_j,
\label{eq:maxsim}
\end{equation}
where $[n]=\{1,\ldots,n\}$.

Given a keep ratio $\rho$, we retain
$k=\lceil \rho n \rceil$ vectors per document. Let $\Phi_{\theta}$ denote a
query-agnostic document-compression function that maps the original document
embeddings to $k$ representative vectors:
\[
R_d
=
\Phi_{\theta}(E_d;\rho)
=
\{\mathbf{r}_1,\ldots,\mathbf{r}_k\}.
\]
The compressed query-document score is computed using the same MaxSim
operation:
\begin{equation}
\widehat{S}_{\theta}(q,d)
=
\sum_{i=1}^{m}
\max_{c \in [k]}
\mathbf{u}_i^{\top}\mathbf{r}_c,
\label{eq:compressed_maxsim}
\end{equation}
where $[k]=\{1,\ldots,k\}$. The compression function $\Phi_{\theta}$ is applied once to each document
during offline indexing. Evaluation queries are not used during compression.

\subsection{Query-Relevant Coverage}
\label{sec:coverage}

For each query embedding $\mathbf{u}_i$, define its winning document patch as:
\[
a_i(q,d)
=
\arg\max_{j\in[n]}
\mathbf{u}_i^\top \mathbf{v}_j,
\]
and let $A(q,d)=\{a_i(q,d)\}_{i=1}^{m}$ denote the winning patches for query
$q$. If compression retains a subset of document-patch indices $P_d\subseteq[n]$, the resulting MaxSim score loss is: 
\begin{equation}
S(q,d)-S_{P_d}(q,d)
=
\sum_{i=1}^{m}
\left[
\mathbf{u}_i^\top\mathbf{v}_{a_i(q,d)}
-
\max_{j\in P_d}
\mathbf{u}_i^\top\mathbf{v}_j
\right].
\label{eq:coverage_loss}
\end{equation}
Retaining all winning patches guarantees exact score preservation, while
approximate preservation requires their responses to be closely reproduced by
retained or synthesized vectors. Because one compressed representation must support many queries, the relevant
document support is the union of $A(q,d)$ over the query distribution.
Compression is easier when this union is concentrated or contains
near-equivalent patches. Thus, covering complementary document regions can be
more useful than selecting individually strong but redundant patches.

For corpus-level analysis, we measure redundancy using
$n/\operatorname{effrank}(E_d)$. Given singular values $\{\sigma_\ell\}$ of the
document embedding matrix, we define
$\operatorname{effrank}(E_d)
=
(\sum_\ell \sigma_\ell^2)^2/\sum_\ell \sigma_\ell^4$.
Larger values indicate more patches per effective embedding direction and
greater potential for substitution. This diagnostic is evaluated in the Discussion under
\emph{Compressibility across content types}; MarginMerge itself optimizes
query-relevant coverage rather than effective rank.

\begin{table*}[t]
\centering
\scriptsize
\setlength{\tabcolsep}{0.5pt}
\renewcommand{\arraystretch}{1.10}
\caption{\textbf{Matched comparison at aggressive vector-retention budgets.}
Mean nDCG@5 at 5\% and 10\% retained document vectors.}
\label{tab:backbone-main}
\begin{adjustbox}{max width=\textwidth,center}
\begin{tabular}{@{}>{\raggedright\arraybackslash}p{4.55cm}
*{7}{>{\centering\arraybackslash}p{0.64cm}}@{\hspace{3pt}}
*{7}{>{\centering\arraybackslash}p{0.64cm}}@{}}
\toprule
\textbf{Method}
& \multicolumn{7}{c}{\cellcolor{mem05}\textbf{5\% vectors retained}}
& \multicolumn{7}{c}{\cellcolor{mem10}\textbf{10\% vectors retained}} \\
\cmidrule(lr){2-8}\cmidrule(lr){9-15}
& \textbf{Arx.} & \textbf{Doc.} & \textbf{Info.} & \textbf{TAT}
& \textbf{Tab} & \textbf{Fli.} & \textbf{Avg.}
& \textbf{Arx.} & \textbf{Doc.} & \textbf{Info.} & \textbf{TAT}
& \textbf{Tab} & \textbf{Fli.} & \textbf{Avg.} \\
\midrule
\rowcolor{bbband}
\multicolumn{15}{c}{\textbf{ColQwen2.5 (Qwen2.5-VL-3B)}~{\tiny \citep{bai2025qwen25vl}}} \\
\midrule
\rowcolor{refgray}
Full index
& .941 & .705 & .928 & .900 & .943 & .932 & .892
& .941 & .705 & .928 & .900 & .943 & .932 & .892 \\
DSE-style~{\tiny \citep{ma2024dse}}
& .640 & .286 & .742 & .481 & .745 & .552 & .574
& .640 & .286 & .742 & .481 & .745 & .552 & .574 \\
\emph{HPC-style}~{\tiny \cite{duong2025hpc}}
& \emph{.884} & \emph{.705} & \emph{.917} & \emph{.897}
& \emph{.903} & \emph{.774} & \emph{.847}
& \emph{.919} & \emph{.703} & \emph{.923} & \emph{.904}
& \emph{.921} & \emph{.894} & \emph{.877} \\
SAP~{\tiny \cite{liu2026look}}
& .880 & .676 & .864 & .864 & .928 & .687 & .817
& .911 & .689 & .892 & .890 & .927 & .799 & .851 \\
Light-ColPali~{\tiny \citep{ma2025storage}}
& .903 & .644 & \textbf{.895} & .883\near & \textbf{.922} & .786 & .839
& .920 & .667 & .909 & .888\near & .927\near & .874 & .864 \\
\rowcolor{oursgray}
\textbf{MarginMerge (ours)}
& \textbf{.927} & \textbf{.688} & .892\near & \textbf{.886}
& .913 & \textbf{.884} & \textbf{.865}
& \textbf{.932} & \textbf{.704} & \textbf{.926} & \textbf{.891}
& \textbf{.930} & \textbf{.916} & \textbf{.883} \\
\midrule
\rowcolor{bbband}
\multicolumn{15}{c}{\textbf{ColPali (PaliGemma-3B)}~{\tiny \citep{faysse2025colpali}}} \\
\midrule
\rowcolor{refgray}
Full index
& .895 & .656 & .874 & .843 & .943 & .956 & .861
& .895 & .656 & .874 & .843 & .943 & .956 & .861 \\
DSE-style~{\tiny \citep{ma2024dse}}
& .730 & .301 & .767 & .416 & .749 & .793 & .626
& .730 & .301 & .767 & .416 & .749 & .793 & .626 \\
\emph{HPC-style}~{\tiny \citep{duong2025hpc}}
& \emph{.801} & \emph{.227} & \emph{.835} & \emph{.835}
& \emph{.875} & \emph{.808} & \emph{.730}
& \emph{.840} & \emph{.437} & \emph{.858} & \emph{.843}
& \emph{.916} & \emph{.888} & \emph{.797} \\
SAP~{\tiny \citep{liu2026look}}
& .845 & .565 & .818 & .682 & .904 & .905 & .787
& .881 & .590 & .852 & .770 & .925 & .933 & .825 \\
Light-ColPali~{\tiny \citep{ma2025storage}}
& .870 & .547 & .859 & .695 & .919 & .944\near & .806
& .887 & .621 & \textbf{.879} & .794 & .934\near & \textbf{.956} & .845 \\
\rowcolor{oursgray}
\textbf{MarginMerge (ours)}
& \textbf{.890} & \textbf{.601} & \textbf{.866} & \textbf{.790}
& \textbf{.940} & \textbf{.946} & \textbf{.839}
& \textbf{.899} & \textbf{.647} & .867 & \textbf{.814}
& \textbf{.938} & .952\near & \textbf{.853} \\
\bottomrule
\end{tabular}
\end{adjustbox}
\end{table*}

\subsection{MarginMerge}
\label{sec:marginmerge}

MarginMerge constructs one synthetic representative for each of $k$ clusters
(Figure~\ref{fig:marginmerge}). It consists of coverage-aware anchor
selection, learned cluster aggregation, and ranking-margin distillation.

\paragraph{Coverage-aware anchors.}

Let $Z=\{\mathbf{z}_t\}_{t=1}^{M}$ be a fixed bank of normalized prototypes
constructed exclusively from training-query embeddings. These prototypes
approximate the query directions that document patches are likely to encounter
at retrieval time. For each prototype $\mathbf{z}_t$, we define its best-match score against the
full patch set of document $d$ as
\[
b_t(d)=\max_{i\in[n]}\mathbf{z}_t^\top\mathbf{v}_i.
\]
We then define the coverage provided by patch $i$ for prototype $t$ as
\begin{equation}
C_{ti}(d)=
\exp\left(
-\frac{b_t(d)-\mathbf{z}_t^\top\mathbf{v}_i}{\tau_c}
\right).
\label{eq:prototype_coverage}
\end{equation}
Because $b_t(d)$ is the maximum response over all patches, the response gap
$b_t(d)-\mathbf{z}_t^\top\mathbf{v}_i$ is non-negative. Thus, values of
$C_{ti}(d)$ near one indicate that patch $i$ closely approximates the
prototype-specific best match, even when it is not the exact maximizing patch.
The temperature $\tau_c$ controls how rapidly coverage decreases with this
response gap.

Given normalized nonnegative prototype weights
$\{\omega_t\}_{t=1}^{M}$, we select $k$ anchors by maximizing
\begin{equation}
F_d(\mathcal{A})=
\sum_{t=1}^{M}
\omega_t
\max_{a\in\mathcal{A}} C_{ta},
\qquad
|\mathcal{A}|=k.
\label{eq:anchor_objective}
\end{equation}
This objective rewards anchor sets that collectively approximate many
training-query directions rather than repeatedly covering the same direction.
In particular, once one selected anchor already provides high coverage for a
prototype, selecting another similar anchor gives little additional gain.

The objective $F_d$ is monotone submodular, so greedy selection obtains the
standard $(1-1/e)$ approximation guarantee. The proof is provided in the
supplementary material.

Let $a_c$ denote the $c$th selected anchor. Each patch is assigned to its
nearest anchor:
\begin{equation}
c(i)=
\arg\max_{c\in[k]}
\mathbf{v}_i^\top\mathbf{v}_{a_c},
\qquad
\mathcal{C}_c=\{i:c(i)=c\}.
\label{eq:cluster_assignment}
\end{equation}
The resulting clusters partition the original document vectors. Anchor
selection is query-distribution-aware, while assignment preserves the local
geometry of the document embedding space.

\paragraph{Learned representative synthesis.}

The prototype bank provides a query-distribution-aware relevance score for
each patch:
\[
s_i(d)=\sum_{t=1}^{M}\omega_t C_{ti}(d).
\]
A patch receives a high score when it closely approximates the strongest
document response for several important prototype directions.

For each patch $i\in\mathcal{C}_c$, we define an unnormalized log-weight
$\ell_i$ that combines anchor similarity, prototype relevance, and a learned
residual:
\begin{equation}
\ell_i
=
\frac{\mathbf{v}_i^\top\mathbf{v}_{a_c}}{\tau_a}
+
p\log\bigl(s_i(d)+\epsilon\bigr)
+
g_\theta(\mathbf{x}_i).
\label{eq:merge_logit}
\end{equation}
Here, $\tau_a$ controls the contribution of anchor similarity, $p$ scales
prototype relevance, and $\epsilon>0$ ensures numerical stability. The small
shared network $g_\theta$ uses compact patch-, anchor-, and cluster-level
features $\mathbf{x}_i$, and its output is clipped to $[-H,H]$. The normalized within-cluster weight and representative are
\begin{align}
\alpha_i
&=
\frac{\exp(\ell_i)}
{\sum_{j\in\mathcal{C}_c}\exp(\ell_j)},
\label{eq:merge_weight}
\\
\mathbf{r}_c
&=
\operatorname{Norm}
\left(
\sum_{i\in\mathcal{C}_c}
\alpha_i\mathbf{v}_i
\right),
\label{eq:representative}
\end{align}
where $\operatorname{Norm}$ denotes $\ell_2$ normalization. Before
normalization, $\mathbf{r}_c$ is a convex combination of the patches assigned
to cluster $c$. Therefore, the representative combines existing document
evidence rather than generating an unconstrained embedding.

\paragraph{Ranking-margin distillation.}

For a training query $q$, let $d^+$ be a relevant document and
$\{d_j^-\}_{j=1}^{N}$ be hard negatives retrieved from the frozen,
uncompressed index. We define the full and compressed positive--negative
margins as
\begin{align}
m_j^{\mathrm{full}}
&=
S(q,d^+)-S(q,d_j^-),
\\
m_j^\theta
&=
\widehat{S}_\theta(q,d^+)
-
\widehat{S}_\theta(q,d_j^-).
\end{align}
We assign larger weights to pairs near the full-index ranking boundary:
\[
\nu_j
=
\frac{
\exp\left(-|m_j^{\mathrm{full}}|/T_m\right)
}{
\sum_{r=1}^{N}
\exp\left(-|m_r^{\mathrm{full}}|/T_m\right)
},
\]
where $T_m$ controls how strongly the weighting concentrates on small-margin
pairs. MarginMerge then minimizes
\begin{equation}
\mathcal{L}_{\mathrm{margin}}
=
\sum_{j=1}^{N}
\nu_j\,
\operatorname{Huber}_{\delta}
\left(
m_j^\theta-m_j^{\mathrm{full}}
\right).
\label{eq:margin_loss}
\end{equation}

Margin matching directly targets the relative score differences that determine
document ordering. A similar compression-induced shift in both positive and
negative scores may change their absolute values while leaving their margin,
and therefore their ordering, nearly unchanged. The Huber loss limits the
influence of unusually large margin errors while remaining quadratic near the
target. The reported full configuration augments this term with auxiliary
ranking and regularization losses described in the supplementary material.
Our factorial study treats the choice of distillation target as an ablation and
does not claim that margin matching is uniformly superior to score reconstruction.

\subsection{Indexing and Retrieval}
\label{sec:indexing}

MarginMerge compresses each document once during offline indexing. The
prototype bank and shared weighting network are used only to construct the
compressed representation $R_d$ and are not required during retrieval.
After indexing, only the $k$ representatives are stored, and queries are scored
using the unchanged MaxSim operation in
Eq.~\eqref{eq:compressed_maxsim}. Thus, MarginMerge requires neither
query-dependent compression nor modification of the retrieval engine.






\begin{table*}[t]
\centering
\scriptsize
\setlength{\tabcolsep}{0.5pt}
\renewcommand{\arraystretch}{1.10}
\caption{\textbf{Selection baselines versus learned representative synthesis.}
nDCG@5 at 5\% and 10\% vector retention. The first four methods retain
original final-layer vectors; MarginMerge synthesizes one vector per cluster.}
\label{tab:backbone-ablation}
\begin{adjustbox}{max width=\textwidth,center}
\begin{tabular}{@{}>{\raggedright\arraybackslash}p{4.55cm}
*{7}{>{\centering\arraybackslash}p{0.64cm}}@{\hspace{3pt}}
*{7}{>{\centering\arraybackslash}p{0.64cm}}@{}}
\toprule
\textbf{Method}
& \multicolumn{7}{c}{\cellcolor{mem05}\textbf{5\% vectors retained}}
& \multicolumn{7}{c}{\cellcolor{mem10}\textbf{10\% vectors retained}} \\
\cmidrule(lr){2-8}\cmidrule(lr){9-15}
& \textbf{Arx.} & \textbf{Doc.} & \textbf{Info.} & \textbf{TAT}
& \textbf{Tab} & \textbf{Fli.} & \textbf{Avg.}
& \textbf{Arx.} & \textbf{Doc.} & \textbf{Info.} & \textbf{TAT}
& \textbf{Tab} & \textbf{Fli.} & \textbf{Avg.} \\
\midrule
\rowcolor{bbband}
\multicolumn{15}{c}
{\textbf{ColQwen2.5 (Qwen2.5-VL-3B)}~{\tiny \citep{bai2025qwen25vl}}}\\
\midrule
Top-likelihood selection
& .669 & .482 & .596 & .614 & .614 & .532 & .585
& .767 & .599 & .740 & .759 & .775 & .724 & .727 \\
Importance-based selection
& .698 & .394 & .817 & .460 & .815 & .364 & .591
& .741 & .469 & .835 & .530 & .854 & .409 & .640 \\
Random selection
& .874 & .609 & .847 & .837 & .865 & .537 & .762
& .907 & .630 & .885 & .870 & .915 & .691 & .816 \\
$k$-center selection
& .826 & .631 & .794 & .811 & .815 & .713 & .765
& .908 & .672 & .884 & .877 & .869 & .841 & .842 \\
\rowcolor{oursgray}
\textbf{MarginMerge (ours)}
& \textbf{.927} & \textbf{.688} & \textbf{.892} & \textbf{.886}
& \textbf{.913} & \textbf{.884} & \textbf{.865}
& \textbf{.932} & \textbf{.704} & \textbf{.926} & \textbf{.891}
& \textbf{.930} & \textbf{.916} & \textbf{.883} \\
\midrule
\rowcolor{bbband}
\multicolumn{15}{c}
{\textbf{ColPali (PaliGemma-3B)}~{\tiny \citep{faysse2025colpali}}}
\\
\midrule
Top-likelihood selection
& .564 & .248 & .400 & .476 & .800 & .873 & .560
& .733 & .414 & .605 & .635 & .855 & .900 & .690 \\
Importance-based selection
& .758 & .386 & .790 & .492 & .792 & .849 & .678
& .781 & .428 & .791 & .533 & .831 & .883 & .708 \\
Random selection
& .827 & .542 & .825 & .623 & .892 & .914 & .771
& .848 & .591 & .856 & .711 & .910 & .937 & .809 \\
$k$-center selection
& .813 & .487 & .719 & .662 & .896 & .892 & .745
& .869 & .555 & .836 & .762 & \textbf{.940} & .929 & .815 \\
\rowcolor{oursgray}
\textbf{MarginMerge (ours)}
& \textbf{.890} & \textbf{.601} & \textbf{.866} & \textbf{.790}
& \textbf{.940} & \textbf{.946} & \textbf{.839}
& \textbf{.899} & \textbf{.647} & \textbf{.867} & \textbf{.814}
& .938\near & \textbf{.952} & \textbf{.853} \\
\bottomrule
\end{tabular}
\end{adjustbox}
\end{table*}

\begin{figure}[t]
\centering
\includegraphics[width=0.98\columnwidth]{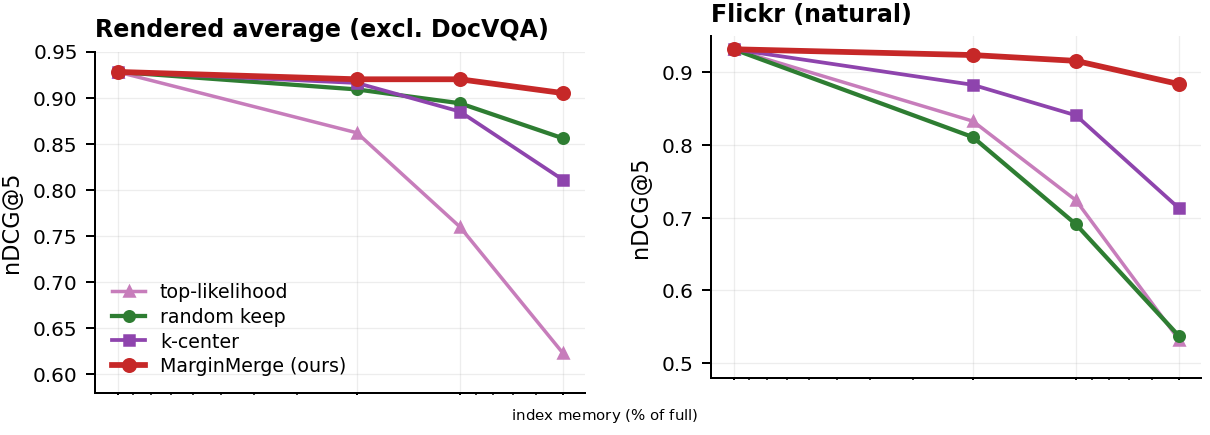}
\caption{\textbf{Retrieval under stronger compression.}
Learned synthesis remains robust as retention falls, especially on Flickr.
Values are reported in Table~\ref{tab:backbone-ablation}.}
\label{fig:curves-main}
\end{figure}

\section{Experiments}
\label{sec:experiments}

\subsection{Experimental Setup}
\label{sec:exp-setup}

\paragraph{Evaluation protocol.}
We compress frozen ColQwen2.5 v0.1 and ColPali representations while keeping
their original MaxSim scorers. We evaluate ArxivQA, DocVQA, InfoVQA, TAT DQA,
TabFQuad, and Flickr. MarginMerge is trained on ArxivQA, TabFQuad, and Flickr.
The remaining document datasets are used only for zero shot evaluation.
COCO, Flickr30k, and TextCaps are used only in the compressibility analysis.
Training and evaluation data are disjoint, as described in Supplementary
Section~A.1. We report nDCG@5 at 5\%, 10\%, and 20\% vector retention and
also measure ranking flips. Most results use three seeds. All methods use the
same corpora, scorers, and full index scores.

\paragraph{Model selection.}
The reported model uses coverage aware anchors, learned representative
synthesis, and the full objective in Supplementary Eq.~(S1). This configuration
is fixed across datasets, backbones, and retention ratios. It is trained at
5\% retention and reused at 10\% and 20\% without retraining. Values within
0.005 of the best matched query agnostic result are marked \near{}.
Significance is discussed only when supported by paired resampling.

\paragraph{Baselines.}
Matched comparisons include Light ColPali geometric merging
\citep{ma2025storage}, a DSE single vector control \citep{ma2024dse}, and
selection controls. The query conditioned HPC component \citep{duong2025hpc}
is included only as an upper reference. SAP \citep{liu2026look} is used
as a diagnostic because its image filtering changes the evaluated subset.
Methods with different backbones or protocols are discussed in Related Work
but are not included in the matched ranking.

\begin{figure}[t]
\centering
\includegraphics[width=\columnwidth]{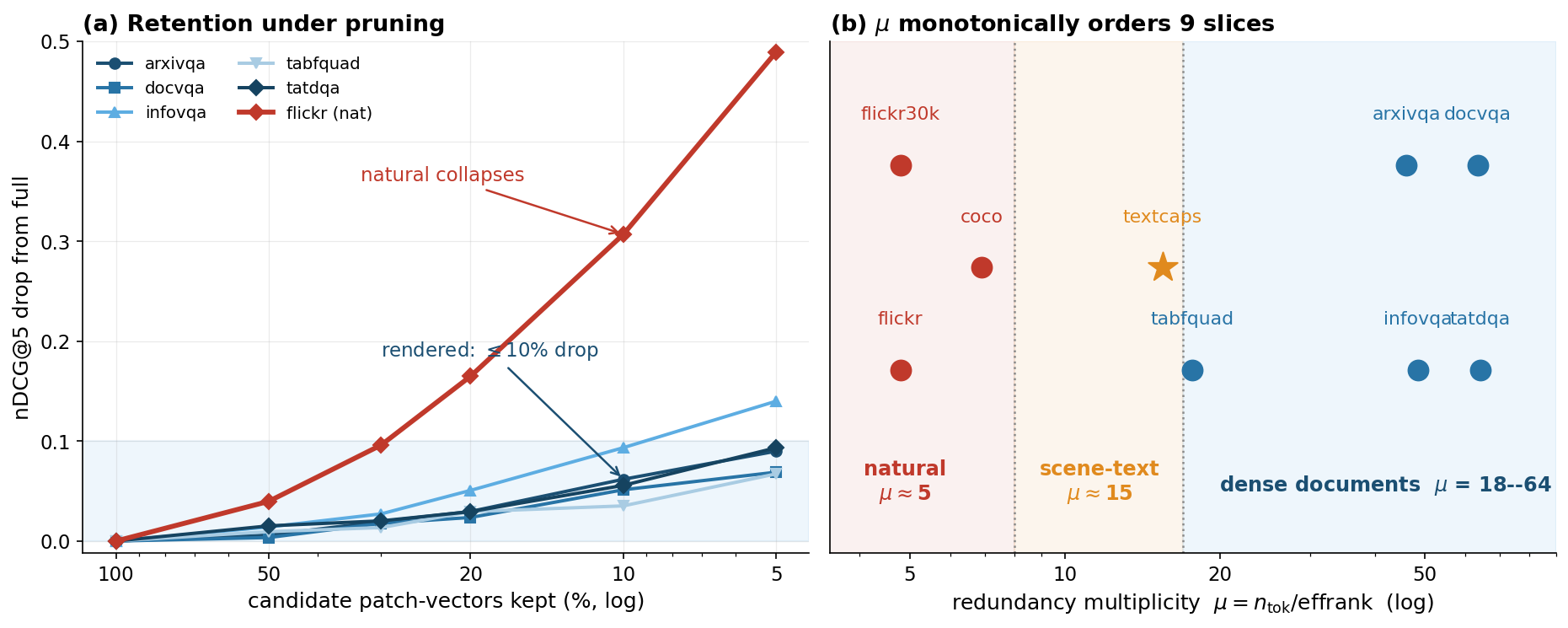}
\caption{\textbf{Compressibility depends on content.}
Random retention is effective for dense rendered pages but degrades on Flickr.
Redundancy multiplicity separates the dataset groups at the corpus level.}
\label{fig:compressibility-main}
\end{figure}

\subsection{Comparison with Compression Baselines}
\label{sec:baseline-comparison}

Table~\ref{tab:backbone-main} shows that MarginMerge achieves the highest
matched query agnostic average at 5\% and 10\% retention on both backbones.
The gains are strongest on Flickr and zero shot DocVQA, while some rendered
document results are close or favor geometric merging. Figure~\ref{fig:curves-main}
shows that the advantage becomes clearer as retention decreases.

\begin{table}[t]
\centering
\scriptsize
\setlength{\tabcolsep}{3.1pt}
\caption{\textbf{Transfer across retention ratios on ColQwen2.5.}
The network trained at 5\% is reused at 10\% and 20\%. Entries are nDCG@5.}
\label{tab:budget-transfer}
\begin{tabular}{lccccccc}
\toprule
Retention & Arx. & Doc. & Info. & TAT & Tab & Fli. & Avg. \\
\midrule
Full index & .941 & .705 & .928 & .900 & .943 & .932 & .892 \\
20\% & .926 & .708 & .928 & .886 & .941 & .924 & .886 \\
10\% & .932 & .704 & .926 & .891 & .930 & .916 & .883 \\
5\%  & .927 & .688 & .892 & .886 & .913 & .884 & .865 \\
\bottomrule
\end{tabular}
\end{table}

\begin{table*}[t]
\centering
\scriptsize
\setlength{\tabcolsep}{0.50pt}
\renewcommand{\arraystretch}{1.04}
\caption{\textbf{Controlled anchor and synthesis factorial across retention
budgets.} ColQwen2.5 nDCG@5 at 5\% and 10\% vector retention. Learned networks
are optimized at 5\% and evaluated at 10\% without retraining.}
\label{tab:factorial-budgets}
\begin{adjustbox}{max width=\textwidth,center}
\begin{tabular}{@{}>{\raggedright\arraybackslash}p{4.55cm}
*{7}{>{\centering\arraybackslash}p{0.64cm}}@{\hspace{3pt}}
*{7}{>{\centering\arraybackslash}p{0.64cm}}@{}}
\toprule
\textbf{Representative / loss}
& \multicolumn{7}{c}{\cellcolor{mem05}\textbf{5\% vectors retained}}
& \multicolumn{7}{c}{\cellcolor{mem10}\textbf{10\% vectors retained}} \\
\cmidrule(lr){2-8}\cmidrule(lr){9-15}
& \textbf{Arx.} & \textbf{Doc.} & \textbf{Info.} & \textbf{TAT}
& \textbf{Tab} & \textbf{Fli.} & \textbf{Avg.}
& \textbf{Arx.} & \textbf{Doc.} & \textbf{Info.} & \textbf{TAT}
& \textbf{Tab} & \textbf{Fli.} & \textbf{Avg.} \\
\midrule
\rowcolor{bbband}
\multicolumn{15}{c}{\textbf{$k$ center anchors}} \\
\midrule
Retained anchor
& .887 & .618 & .795 & .867 & .833 & .626 & .771
& .908 & .661 & .862 & .886 & .914 & .756 & .831 \\
Uniform centroid
& .924 & .650 & .899 & .893 & .923 & .857 & .857
& .933 & .677 & .917 & .899 & .937 & .913 & .879 \\
Response centroid
& .918 & .666 & .848 & .891 & .864 & .719 & .818
& .935 & .692 & .896 & .901 & .925 & .838 & .864 \\
Learned, score recon.
& .928 & .708 & .908 & .896 & .935 & .891 & \textbf{.877}
& .937 & .704 & .921 & .900 & .944 & .921 & \textbf{.887} \\
Learned, margin
& .931 & .697 & .910 & .894 & .929 & .888 & .875
& .936 & .706 & .918 & .899 & .939 & .918 & .885 \\
Learned, full
& .931 & .693 & .911 & .894 & .930 & .891 & .875
& .935 & .707 & .918 & .901 & .942 & .918 & .886 \\
\midrule
\rowcolor{bbband}
\multicolumn{15}{c}{\textbf{Coverage aware anchors}} \\
\midrule
Retained anchor
& .921 & .631 & .871 & .856 & .883 & .759 & .820
& .923 & .646 & .885 & .857 & .931 & .872 & .852 \\
Uniform centroid
& .937 & .669 & .893 & .895 & .915 & .821 & .854
& .939 & .720 & .933 & .899 & .940 & .902 & .889 \\
Response centroid
& .925 & .652 & .877 & .878 & .889 & .810 & .838
& .928 & .678 & .899 & .885 & .931 & .900 & .871 \\
Learned, score recon.
& .931 & .693 & .899 & .898 & .931 & .892 & .875
& .939 & .711 & .939 & .904 & .942 & .926 & \textbf{.894} \\
Learned, margin
& .938 & .699 & .899 & .897 & .924 & .895 & \textbf{.876}
& .941 & .710 & .935 & .901 & .940 & .927 & .893 \\
\rowcolor{oursgray}
Learned, full
& .927 & .688 & .892 & .886 & .913 & .884 & .865
& .932 & .704 & .926 & .891 & .930 & .916 & .883 \\
\bottomrule
\end{tabular}
\end{adjustbox}
\end{table*}

\paragraph{Stability and transfer.}
At 5\% retention, MarginMerge reduces ranking flips relative to geometric
merging on all six ColQwen2.5 datasets, with an average reduction of about
41\%. Seed results and flip rates are reported in Supplementary Table~S1.
The zero shot datasets are excluded from all fitting stages. The leave one
dataset out study also improves over the fixed response centroid on every held
out target, as shown in Supplementary Table~S2.




\subsection{Ablation of Design Choices}
\label{sec:ablations}

Table~\ref{tab:backbone-ablation} compares MarginMerge with several patch
selection strategies on both ColQwen2.5 and ColPali. Random and $k$ center
selection are stronger than salience based alternatives, which supports the
need to preserve diverse document regions. MarginMerge performs best on
average across both backbones and retention budgets, showing that constructing
new representatives is more effective than retaining a subset of the original
patch vectors.

Table~\ref{tab:factorial-budgets} further separates the effects of anchor
selection, representative construction, and training objective. Coverage aware
anchors provide their clearest benefit when the representative is fixed, while
learned synthesis produces the largest overall improvement and reduces
sensitivity to the anchor rule. Score reconstruction and margin matching
perform similarly, indicating that learning how patches are combined is more
important than the precise distillation objective. The reported full
configuration is fixed in advance and used for all main comparisons, transfer
experiments, and stability analyses rather than selected separately for each
setting.

\begin{figure*}[t]
\centering
\begin{minipage}{0.315\textwidth}
\centering
\includegraphics[width=0.90\linewidth]{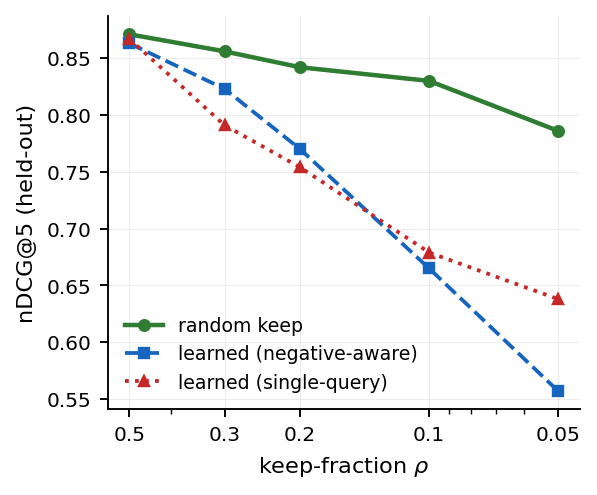}
{\small \textbf{(a)} Learned pruning}
\end{minipage}\hfill
\begin{minipage}{0.315\textwidth}
\centering
\includegraphics[width=0.90\linewidth]{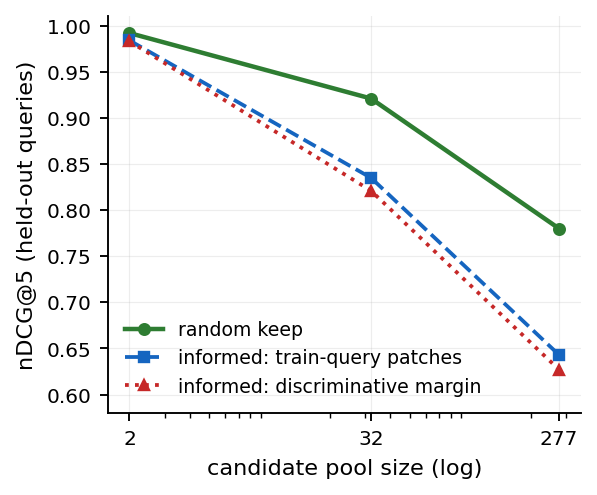}
{\small \textbf{(b)} Candidate pool scaling}
\end{minipage}\hfill
\begin{minipage}{0.315\textwidth}
\centering
\includegraphics[width=0.90\linewidth]{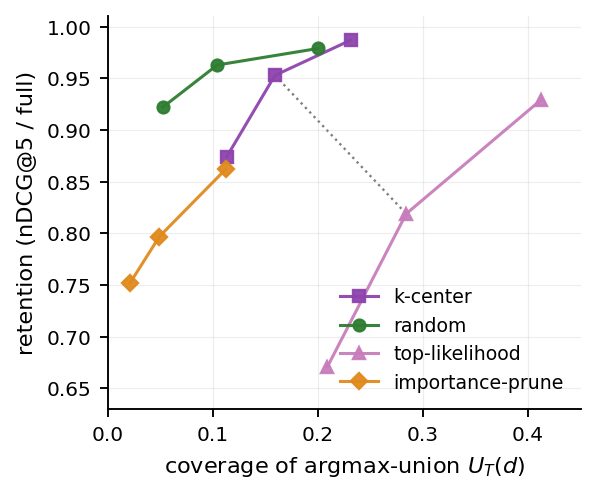}
{\small \textbf{(c)} Winner coverage versus retention}
\end{minipage}
\caption{\textbf{Why query agnostic pruning fails.}
Learned selectors can underperform random retention on unseen queries.
The gap grows with the candidate pool, and diversity preserves retrieval better
than selecting patches only by predicted likelihood.}
\label{fig:pruning-main}
\end{figure*}

\subsection{Discussion}
\label{sec:analysis}

\paragraph{Compressibility across content types.}
Figure~\ref{fig:compressibility-main} supports the coverage account at the
corpus level. Natural images have fewer substitutable patch directions, while
dense rendered documents contain greater redundancy. The same ordering holds across nine datasets and under a fixed-patch-count
control in Supplementary Section B.2. Individual document statistics remain weak predictors, so we use
redundancy as a dataset level explanation rather than a causal measure.

\paragraph{Limitations of Query Agnostic Pruning.}
Figure~\ref{fig:pruning-main} shows that salience alone does not preserve
the evidence needed by unseen queries. Learned selectors fit their training
queries but may retain similar patches and miss complementary regions.
Diversity based selection preserves retrieval more reliably, even when it
covers fewer observed winners. A compact reusable index must therefore
represent diverse query directions rather than only the patches favored by
seen queries.

\subsection{Efficiency, Scope, and Limitations}
\label{sec:scope}

MarginMerge reduces stored vectors and exact MaxSim comparisons in proportion
to the retention ratio and requires no additional model execution during
retrieval. Greedy anchor selection remains expensive during indexing, taking
about 1.7 seconds per TAT DQA document in our implementation. We report vector
retention rather than storage in bytes and do not compare complete search
latency with quantized retrieval systems. The redundancy analysis is
observational. Faster anchor selection, byte level evaluation, and controlled
redundancy studies are important directions for future work.


\section{Conclusion}
We studied compression of frozen multi vector visual document retrievers with
a focus on query relevant coverage. MarginMerge constructs compact synthetic
representatives offline while preserving the standard online MaxSim interface.
Across six datasets, two backbones, and aggressive vector retention budgets, it
improves the balance between retrieval quality and storage cost and reduces
ranking changes caused by compression. Controlled ablations show that learned
representative synthesis is the dominant source of improvement. Coverage based
anchors help most when aggregation is constrained, while learned weighting
reduces sensitivity to the anchor rule. Score reconstruction and margin
matching perform similarly, and the auxiliary terms in the reported full
objective do not account for its mean nDCG gains. These findings refine the
coverage account and motivate faster coverage initialization and more
transferable representative synthesis.
\bibliography{aaai2027}

\appendix
\setcounter{secnumdepth}{2}
\setcounter{table}{0}
\renewcommand{\thetable}{S\arabic{table}}
\setcounter{figure}{0}
\renewcommand{\thefigure}{S\arabic{figure}}
\setcounter{equation}{0}
\renewcommand{\theequation}{S\arabic{equation}}

\section{Experimental Protocol and Implementation}
\label{app:experimental-details}

\subsection{Data Splits and Leakage Controls}
\label{app:protocol}
\label{app:leakage}

All main compression results use the same full corpus evaluation manifests and
full index scores. The manifests are shared by MarginMerge, geometric merging,
the full index reference, and all retention calculations. DocVQA is excluded
only from the rendered document aggregate in the main retrieval and
compressibility figures. Its results remain in every per dataset comparison
table. For each retrieval dataset, 30\% of the document identifiers are selected with
\textnormal{RandomState(0)} and all queries associated with these documents form the
evaluation set. The remaining documents form the training side of the split.
Training and evaluation documents are disjoint by identifier. A separate audit
also compares SHA256 hashes of the frozen patch embeddings and finds no content
overlap between the evaluation documents and the documents used for training,
validation, negative mining, feature normalization, or cluster construction.

The weighting network is optimized on ArxivQA, TabFQuad, and Flickr. Within
these source datasets, 10\% of the training side queries are assigned to
validation with split seed 42. Evaluation queries are excluded from prototype
construction, hard negative mining, feature normalization, training,
validation, and checkpoint selection. The global prototype bank uses training
side query tokens from all six retrieval datasets. Thus, DocVQA, InfoVQA, and
TAT-DQA are unseen by the weighting network, while their training side query
tokens contribute to the shared prototype bank. In the leave one dataset out
study, the target dataset is also removed from prototype construction,
normalization, negative mining, training, validation, and checkpoint selection.
The evaluations at 10\% and 20\% reuse the network optimized at 5\% vector
retention without further training.

A final audit identified document overlap in an earlier enlarged TabFQuad
diagnostic manifest. No result from that manifest is used in the main paper or
in the supplementary claims. The loss component analysis uses only the
nonoverlapping ArxivQA and Flickr diagnostics.

\subsection{Datasets and Frozen Encoders}
\label{app:data-encoding}

{\small\raggedright
The retrieval experiments use the public test splits
{\normalfont vidore/\hspace{0pt}arxivqa\_\hspace{0pt}test\_\hspace{0pt}subsampled},
{\normalfont vidore/\hspace{0pt}docvqa\_\hspace{0pt}test\_\hspace{0pt}subsampled},
{\normalfont vidore/\hspace{0pt}infovqa\_\hspace{0pt}test\_\hspace{0pt}subsampled},
{\normalfont vidore/\hspace{0pt}tatdqa\_\hspace{0pt}test},
{\normalfont vidore/\hspace{0pt}tabfquad\_\hspace{0pt}test\_\hspace{0pt}subsampled}, and
{\normalfont nlphuji/\hspace{0pt}flickr\_\hspace{0pt}1k\_\hspace{0pt}test\_\hspace{0pt}image\_\hspace{0pt}text\_\hspace{0pt}retrieval}. TAT-DQA and
TabFQuad pages are deduplicated by \textnormal{image\_filename}. Queries that are
null or empty are removed. The compressibility analysis additionally uses the
test splits of {\normalfont nlphuji/\hspace{0pt}mscoco\_\hspace{0pt}2014\_\hspace{0pt}5k\_\hspace{0pt}test\_\hspace{0pt}image\_\hspace{0pt}text\_\hspace{0pt}retrieval}
and {\normalfont nlphuji/\hspace{0pt}flickr30k}, together with the validation split of
{\normalfont lmms-lab/\hspace{0pt}TextCaps}.
\par}

All images are converted to RGB and passed to the original model processor.
No manual resizing or crop rule is added. ColQwen2.5 uses image batches of four
for document pages and sixteen for Flickr. ColPali uses image batches of eight.
Both encoders use query batches of 32. The frozen models are loaded in bfloat16,
and the masked query and document vectors are cached in float16. All MaxSim
scores use only nonpadding vectors. The two frozen checkpoints are
\textnormal{vidore/colqwen2.5-v0.1} and \textnormal{vidore/colpali-v1.3}.

\subsection{Metrics and Statistical Reporting}
\label{app:metrics}

For a document with $n$ patch vectors and retention ratio $\rho$, every method
stores
\[
 k=\lceil \rho n\rceil
\]
vectors unless the method has a fixed single vector representation. Retrieval
quality is measured with nDCG@5. For query $q$, we compute
\[
\begin{aligned}
\operatorname{DCG@5}(q)
&=
\sum_{r=1}^{5}
\frac{2^{\operatorname{rel}_{q,r}}-1}{\log_2(r+1)},\\
\operatorname{nDCG@5}(q)
&=
\frac{\operatorname{DCG@5}(q)}{\operatorname{IDCG@5}(q)}.
\end{aligned}
\]
The reported score is the mean across evaluation queries. Each query has one
relevant page in the evaluation protocol, so the score is determined by the
rank of that page within the first five results.

For a relevant document $d^+$ and a negative document $d^-$, define the full
and compressed margins as
\[
\begin{aligned}
m^{\mathrm{full}}&=S(q,d^+)-S(q,d^-),\\
m^\theta&=\widehat S_\theta(q,d^+)-\widehat S_\theta(q,d^-).
\end{aligned}
\]
A compression induced ranking flip occurs when
$\operatorname{sign}(m^{\mathrm{full}})\ne\operatorname{sign}(m^\theta)$.
The flip rate is the fraction of evaluated pairs that satisfy this condition.

The three seed results use seeds 42, 43, and 44. The document split uses seed 0
and the training and validation split uses seed 42 for every run. Table
\ref{tab:app-mm-seeds} reports the individual seed values, mean, and standard
deviation. ColPali transfer slices and diagnostic experiments use seed 42.
Values within 0.005 of the best matched query agnostic result are marked as
numerically close in the main paper. This marker is not a statistical
significance claim. The factorial analysis uses paired bootstrap intervals over per query nDCG@5
differences. Matching seed and query observations are paired, then 2,000
bootstrap samples are drawn with replacement using \textnormal{RandomState(0)}.
The reported interval uses the 2.5 and 97.5 percentiles. An improvement is
marked as supported only when the paired interval excludes zero and the
absolute mean difference is at least 0.005.

\subsection{Prototype Bank and Weighting Network}
\label{app:prototype-network}

The prototype bank contains $M=128$ normalized query directions. At most 32
query tokens are sampled from each training side query, with equal total weight
assigned to every query. The pooled token set is capped at one million tokens.
The bank is constructed with spherical clustering, implemented by clustering
normalized vectors and normalizing each centroid. MiniBatchKMeans is used when
the token count exceeds 20,000, with batch size 8,192, 50 iterations, five
initializations, and seed 42. Full KMeans uses the same iteration count,
initialization count, and seed for smaller token sets. If $f_t$ is the weighted
frequency assigned to prototype $t$, its normalized weight is
\[
\omega_t=
\frac{(f_t+10^{-8})^{1/2}}
{\sum_{r=1}^{M}(f_r+10^{-8})^{1/2}}.
\]
Coverage uses $\tau_c=0.05$. The base cluster weight uses $\tau_a=0.1$,
$p=0.5$, and $\epsilon=10^{-8}$. The resulting unnormalized weight is
\[
\widetilde\alpha_i=
\exp\!\left(\frac{v_i^\top v_{a_c}}{0.1}\right)
\left(s_i+10^{-8}\right)^{1/2}\exp(h_i),
\]
where $h_i$ is the learned residual. Weights are normalized within each cluster
before representative synthesis. The shared network uses fifteen scalar features for each patch. They are anchor
similarity, similarity to the uniform centroid, similarity to the response
centroid, prototype response importance, log cluster size, relative cluster
size, within cluster distance rank, two spatial coordinates, an anchor
indicator, local redundancy, weighted mean prototype response, maximum
prototype response, weighted response standard deviation, and weighted
coverage. Patch grid metadata is not stored in the cached embeddings, so the
two spatial coordinates are set to zero in all reported runs. Features are
standardized with means and standard deviations computed only from training
side documents.

The residual network has dimensions $15\rightarrow32\rightarrow16\rightarrow1$
and uses GELU after both hidden layers. It contains 1,057 trainable parameters.
The output is clipped to $[-5,5]$, which sets $H=5$. The encoders, prototype
bank, anchors, assignments, and base features remain frozen.

\begin{table*}[t]
\centering
\scriptsize
\setlength{\tabcolsep}{3.8pt}
\caption{\textbf{Per seed results at 5\% vector retention.}
$\Delta$ is the gain over
geometric merging from unrounded values.}
\label{tab:app-mm-seeds}
\begin{tabular}{lccccccc}
\toprule
Dataset & Seed 42 & Seed 43 & Seed 44 & Mean $\pm$ std & Merging & $\Delta$ & Flip \\
\midrule
ArxivQA & 0.9245 & 0.9246 & 0.9324 & $0.9272\pm0.0037$ & 0.903 & $+0.024$ & .0033/.0048 \\
DocVQA & 0.6876 & 0.6825 & 0.6944 & $0.6882\pm0.0049$ & 0.644 & $+0.044$ & .0621/.1095 \\
InfoVQA & 0.8945 & 0.8907 & 0.8902 & $0.8918\pm0.0019$ & 0.895 & $-0.003$ & .0080/.0097 \\
TAT-DQA & 0.8874 & 0.8845 & 0.8874 & $0.8864\pm0.0014$ & 0.883 & $+0.003$ & .0082/.0118 \\
TabFQuad & 0.9160 & 0.9064 & 0.9160 & $0.9128\pm0.0045$ & 0.922 & $-0.009$ & .0093/.0113 \\
Flickr & 0.8847 & 0.8767 & 0.8901 & $0.8838\pm0.0055$ & 0.786 & $+0.097$ & .0031/.0132 \\
\bottomrule
\end{tabular}
\end{table*}

\begin{table}[t]
\centering
\scriptsize
\setlength{\tabcolsep}{3.0pt}
\caption{\textbf{Leave one dataset out transfer at 5\% vector retention.}}
\label{tab:app-loo}
\begin{tabular}{lcccc}
\toprule
Target & Merging & Centroid & MarginMerge & Gain \\
\midrule
Flickr & 0.7536 & 0.6692 & \textbf{0.6840} & $+0.0148$ \\
TabFQuad & 0.8488 & 0.8233 & \textbf{0.8573} & $+0.0340$ \\
ArxivQA & 0.8845 & 0.9245 & \textbf{0.9334} & $+0.0089$ \\
\bottomrule
\end{tabular}
\end{table}

\begin{table}[h!]
\centering
\small
\setlength{\tabcolsep}{3.2pt}
\caption{\textbf{Redundancy multiplicity across nine datasets.}}
\label{tab:app-redundancy}
\begin{tabular}{llccc}
\toprule
Dataset & Type & \#vectors & $\mu$ & Ret.@10\% \\
\midrule
Flickr30k & natural & 237  & 4.8  & NA \\
Flickr    & natural & 238  & 4.8  & 65\% \\
COCO      & natural & 361  & 6.9  & NA \\
TextCaps  & scene text & 1011 & 15.5 & NA \\
TabFQuad  & dense & 1175 & 17.7 & 96\% \\
ArxivQA   & dense & 2927 & 46.1 & 93\% \\
InfoVQA   & dense & 3837 & 48.5 & 90\% \\
DocVQA    & dense & 4552 & 63.4 & 91\% \\
TAT-DQA   & dense & 4862 & 64.1 & 93\% \\
\bottomrule
\end{tabular}
\end{table}

\begin{table}[h!]
\centering
\scriptsize
\setlength{\tabcolsep}{2.7pt}
\caption{\textbf{Per instance density proxies and pruning sensitivity.}}
\label{tab:app-proxies}
\begin{tabular}{lcccc}
\toprule
Dataset & \#patch & Eff. rank & Redund. & \#q-tok \\
\midrule
ArxivQA & $-0.10$ & $+0.01$ & $-0.19$ & $-0.19$ \\
DocVQA  & $-0.03$ & $+0.09$ & $-0.11$ & $+0.01$ \\
InfoVQA & $-0.21$ & $+0.01$ & $-0.23$ & $-0.16$ \\
Flickr  & $+0.02$ & $+0.15$ & $-0.20$ & $+0.05$ \\
\bottomrule
\end{tabular}
\end{table}

\begin{table}[h!]
\centering
\scriptsize
\setlength{\tabcolsep}{2.7pt}
\caption{\textbf{Distinctiveness salience and random retention.}}
\label{tab:app-salience}
\begin{tabular}{llccccc}
\toprule
Dataset & Method & 50\% & 30\% & 20\% & 10\% & 5\% \\
\midrule
\multirow{2}{*}{ArxivQA}
 & Random   & .871 & .857 & .847 & .827 & .801 \\
 & Salience & .841 & .807 & .791 & .735 & .638 \\
\multirow{2}{*}{InfoVQA}
 & Random   & .910 & .892 & .881 & .837 & .791 \\
 & Salience & .863 & .829 & .765 & .680 & .561 \\
\multirow{2}{*}{DocVQA}
 & Random   & .564 & .566 & .556 & .531 & .500 \\
 & Salience & .542 & .492 & .479 & .443 & .376 \\
\multirow{2}{*}{Flickr}
 & Random   & .832 & .769 & .694 & .564 & .397 \\
 & Salience & .749 & .620 & .502 & .305 & .179 \\
\bottomrule
\end{tabular}
\end{table}

\begin{table}[h!]
\centering
\small
\caption{\textbf{Learned and random pruning on ArxivQA.}}
\label{tab:app-learned}
\begin{tabular}{cccc}
\toprule
Keep ratio & Single query & Negative aware & Random \\
\midrule
.50 & .867 & .863 & \textbf{.871} \\
.30 & .791 & .823 & \textbf{.856} \\
.20 & .754 & .770 & \textbf{.842} \\
.10 & .679 & .665 & \textbf{.830} \\
.05 & .638 & .557 & \textbf{.786} \\
\bottomrule
\end{tabular}
\end{table}

\begin{table}[h!]
\centering
\scriptsize
\setlength{\tabcolsep}{2.8pt}
\caption{\textbf{Empirical winner coverage and retrieval retention.}}
\label{tab:app-raw-coverage}
\begin{tabular}{lcccc}
\toprule
& \multicolumn{2}{c}{10\% vectors} & \multicolumn{2}{c}{5\% vectors} \\
Method & Coverage & Retention & Coverage & Retention \\
\midrule
Random & .104 & .963 & .052 & .922 \\
Importance pruning & .049 & .797 & .021 & .752 \\
Top likelihood & \textbf{.284} & .819 & \textbf{.209} & .671 \\
$k$-center & .159 & \textbf{.953} & .113 & \textbf{.874} \\
\bottomrule
\end{tabular}
\end{table}

\begin{table*}[h!]
\centering
\scriptsize
\setlength{\tabcolsep}{5.0pt}
\caption{\textbf{Overlap free loss component diagnostic at 5\% vector
retention.}}
\label{tab:app-loss-ablation}
\begin{tabular}{lcc}
\toprule
Variant & Flickr & ArxivQA \\
\midrule
Fixed response centroid & .0000/.0078/.0074 & .0000/.0046/.0038 \\
Full objective & \textbf{+.0780}/.0041/.0037 & +.0120/.0036/.0027 \\
Without listwise loss & +.0764/.0042/.0038 & +.0083/.0038/.0027 \\
Without boundary weighting & +.0727/.0042/.0039 & \textbf{+.0159}/.0036/.0026 \\
Margin only & +.0714/.0044/.0041 & \textbf{+.0159}/.0038/.0027 \\
\bottomrule
\end{tabular}
\end{table*}

\subsection{Training Objective and Optimization}
\label{app:mm-training}
\label{app:cost}

The complete training objective augments the pairwise margin loss with ranking
and regularization terms.
\begin{equation}
\begin{aligned}
\mathcal{L}={}&
\lambda_m\mathcal{L}_{\mathrm{margin}}
+\lambda_r\mathcal{L}_{\mathrm{rank}}
+\lambda_l\mathcal{L}_{\mathrm{list}} \\
&+\lambda_w\mathcal{L}_{\mathrm{weight}}
+\lambda_e\mathcal{L}_{\mathrm{entropy}}
+\lambda_a\mathcal{L}_{\mathrm{anchor}}.
\end{aligned}
\label{eq:app-full-loss}
\end{equation}
The coefficients are $\lambda_m=1$, $\lambda_r=0.5$, $\lambda_l=0.5$,
$\lambda_w=10^{-3}$, $\lambda_e=0.01$, and $\lambda_a=0.01$. The margin loss
uses eight hard negatives, $T_m=2$, and Huber threshold $\delta=0.5$. Hard
negatives are the eight highest scoring nonrelevant training documents under
the frozen full index.

Let $\widehat m_j$ denote a compressed positive and negative margin. The ranking
term uses a margin threshold of $\gamma=0.2$.
\[
\mathcal{L}_{\mathrm{rank}}
=
\frac{1}{N}\sum_{j=1}^{N}\max(0,\gamma-\widehat m_j).
\]
Let $\mathbf{s}$ and $\widehat{\mathbf{s}}$ contain the full and compressed
scores for the positive document followed by its negatives. The listwise term
is
\[
\mathcal{L}_{\mathrm{list}}
=
D_{\mathrm{KL}}\!\left(
\operatorname{softmax}(\mathbf{s})
\,\middle\|\,
\operatorname{softmax}(\widehat{\mathbf{s}})
\right).
\]
The residual penalty is the mean squared log weight
\[
\mathcal{L}_{\mathrm{weight}}=\frac{1}{P}\sum_{i=1}^{P}h_i^2,
\]
where $P$ counts the patches in the positive and negative documents used for
that training example.

For cluster $c$, define the normalized entropy
\[
E_c=
\frac{-\sum_{i\in C_c}\alpha_i\log(\alpha_i+10^{-8})}
{\log(\max(|C_c|,2))}.
\]
The entropy term averages $\max(0,0.2-E_c)^2$ across clusters and documents.
The anchor term averages
\[
\max\!\left(0,\frac{0.25}{|C_c|}-\alpha_{a_c}\right)^2
\]
across clusters and documents. These two terms discourage collapsed cluster
weights and vanishing anchor contributions. Table
\ref{tab:app-loss-ablation} shows that the auxiliary terms have a small effect
on held out nDCG@5 relative to the effect of learned representative synthesis.

The network is optimized with AdamW using learning rate $3\times10^{-4}$ and
weight decay $10^{-4}$. Gradients are accumulated across four queries, and the
gradient norm is clipped to 1.0 before every update. Training runs for at most
five epochs. It stops after two epochs without validation improvement. The
selected checkpoint has the highest validation nDCG@5, with lower validation
flip rate used to resolve a tie. The same scalar settings are used for every
dataset, backbone, retention ratio, and seed. The archived development protocol compares three anchor rules, four
representative constructions, and three objectives for the learned
representative. A preliminary phase evaluates ArxivQA, Flickr, and TabFQuad at
5\% retention with seed 42. The confirmatory phase evaluates all six datasets
at 5\% and 10\% retention with seeds 42, 43, and 44. These seeds define three
independent repetitions. The network trained at 5\% retention is reused at
10\% and 20\%. The SAP reproduction separately evaluates layers
$\{14,18,19,21,25\}$ on validation data and fixes layer 25 for every reported
dataset. No scalar setting or checkpoint is selected separately for an
evaluation dataset.

For $M$ query prototypes, $n$ document vectors, and $k$ representatives, exact
greedy anchor selection requires $\mathcal{O}(Mnk)$ similarity updates.
Cluster assignment requires $\mathcal{O}(nkD)$. The shared weighting network
adds 1 to 4\,ms per document after the anchors are available. Construction is
dominated by anchor selection, which takes approximately 1.7\,s per TAT-DQA
document in the reported implementation. This cost is paid once during
indexing. Retrieval stores only the $k$ representatives and uses the original
MaxSim engine.

\subsection{Computing Environment}
\label{app:environment}

All experiments were run on one NVIDIA B200 GPU with 256\,GB of GPU memory.
The host had 256\,GB of system memory. The ColQwen environment used PyTorch
2.8.0, the CUDA 12.8 runtime, datasets 3.6.0, NumPy 2.4.2, scikit-learn 1.8.0,
pandas 3.0.3, and torchvision 0.23.0. ColQwen encoding used transformers
4.47.1. Cached embedding experiments and analysis used transformers 4.55.0.
The ColPali environment used PyTorch 2.13.0, the CUDA 13.0 runtime,
transformers 4.46.3, colpali-engine 0.3.5, datasets 5.0.0, peft 0.11.1, and
NumPy 1.26.4. CPU thread counts were restricted during clustering and analysis
to avoid oversubscription. The submitted code archive contains the complete
package freezes for both environments. Code, preprocessing, evaluation, and
analysis scripts will be released publicly upon publication under terms that
permit academic research use.

\subsection{SAP Reproduction}
\label{app:sap}

SAP uses a middle layer anchor score for structural relevance. We sweep layers
$\{14,18,19,21,25\}$, which correspond to approximately 39\% to 69\% of the
network depth, on two validation slices. Layer 25 is selected once and then
used for every reported dataset. Layer 19 falls below its matched random
baseline on ArxivQA. Layer 25 exceeds the matched random baseline on both
validation datasets. Oversized images exceed eager attention memory on parts of InfoVQA, TAT-DQA,
and DocVQA. The retained coverage is 92.7\%, 97.6\%, and 98.7\%, respectively.
All other SAP datasets have complete coverage. Since these filtered subsets do
not exactly match the subsets used by the other methods, SAP is treated as a
diagnostic and is excluded from aggregate rankings.
\section{Additional Results}
\label{app:additional-results}

\subsection{Stability and Cross Dataset Transfer}
\label{app:mm-perseed}

The seed standard deviations are below 0.006 on all six datasets. The largest
gains over geometric merging occur on Flickr, DocVQA, and ArxivQA. In the leave
one dataset out study, learned weighting improves over the fixed response
centroid on every held out target. The Flickr result remains below geometric
merging, so the transfer result supports learned weighting rather than uniform
dominance over every baseline.

\subsection{Compressibility Controls}
\label{app:compressibility}

At a fixed absolute budget of $k=32$, TabFQuad, InfoVQA, and Flickr obtain
nDCG@5 scores of 0.797, 0.718, and 0.622. After InfoVQA is subsampled to the
same input size as Flickr, approximately 238 vectors, its score remains 0.718.
Raw patch count therefore does not explain the difference between rendered
content and natural images. The weak correlations in
Table~\ref{tab:app-proxies} also show that corpus level redundancy should not be
used as a document level confidence score.

\subsection{Pruning Diagnostics}
\label{app:pruning}

The salience rule falls below random retention at every reported setting in
Table~\ref{tab:app-salience}. The two learned selectors also fall below random
on held out ArxivQA, with a wider gap at lower retention. The candidate pool
experiment in the main paper reaches the same conclusion at pool sizes 2, 32,
and 277. Table~\ref{tab:app-raw-coverage} separates raw winner count from diverse
coverage. Top likelihood keeps the largest fraction of observed winners, but
$k$-center retains more retrieval quality. The result supports the claim that a
query agnostic index must represent complementary directions rather than
concentrating on similar winners. The negative aware selector uses LoRA on frozen ColQwen2.5, a soft gated
approximation to MaxSim, a budget regularizer, and gradient clipping. An early
straight through variant added $(\mathrm{keep}-1)\times10^{4}$ to the forward
mask, which produced gradients at the same scale and unstable optimization.
All reported results use the soft gate formulation. The discarded formulation
is not used in any table or claim.

\subsection{Loss Component Ablation}
\label{app:loss-ablation}

All learned variants improve over the fixed response centroid on both
diagnostics. Removing listwise distillation, boundary weighting, or the
auxiliary ranking term changes the gain by at most 0.007 on Flickr and 0.004 on
ArxivQA. The flip and reversal rates are also similar across the learned
variants. These results agree with the main factorial study. Learned
representative synthesis provides the consistent gain, while the precise
auxiliary loss composition has a smaller effect. Margin matching is not claimed
to be uniformly better than score reconstruction.

\section{Proofs}
\label{app:proofs}

\subsection{Coverage Anchor Objective}
\label{app:coverage-proof}

For a fixed prototype $t$, define
$f_t(\mathcal{A})=\max_{a\in\mathcal{A}}C_{ta}$ and
$f_t(\emptyset)=0$. For $\mathcal{A}\subseteq\mathcal{B}$ and
$x\notin\mathcal{B}$,
\begin{align}
&f_t(\mathcal{A}\cup\{x\})-f_t(\mathcal{A}) \\
&\qquad=\max\{0,C_{tx}-f_t(\mathcal{A})\}.
\end{align}
Since $f_t(\mathcal{A})\le f_t(\mathcal{B})$, the marginal gain of adding
$x$ to $\mathcal{A}$ is at least its marginal gain when added to
$\mathcal{B}$. Thus, $f_t$ is monotone and submodular. Nonnegative weighted
sums preserve both properties, so
$F_d(\mathcal{A})=\sum_t\omega_t f_t(\mathcal{A})$ is monotone and
submodular. The greedy algorithm therefore obtains the standard
$(1-1/e)$ approximation under a cardinality constraint.

\subsection{Sufficient Condition for No Ranking Flip}
\label{app:no-flip-proof}

Let
$\epsilon_\theta(q,d)=\widehat{S}_\theta(q,d)-S(q,d)$ and define
\begin{equation}
\Delta_j=\epsilon_\theta(q,d^+)-\epsilon_\theta(q,d_j^-).
\end{equation}
The compressed margin satisfies
\begin{equation}
m_j^\theta=m_j^{\mathrm{full}}+\Delta_j.
\end{equation}
If $|\Delta_j|<|m_j^{\mathrm{full}}|$, the perturbation cannot move the margin
across zero. It follows that
$\operatorname{sign}(m_j^\theta)=\operatorname{sign}(m_j^{\mathrm{full}})$,
and the ordering is preserved.

\end{document}